# CASE STUDY ON SOCIAL ENGINEERING TECHNIQUES FOR PERSUASION

Mosin Hasan[1], Nilesh Prajapati[2] and Safvan Vohara[3]

[1]Computer Department, BVM Engineering College, V V Nagar
mosin83@yahoo.co.in
[2IT] Department, BVM Engineering College, V V Nagar
nbp_it53@yahoo.com
[3]Computer Department, BVM Engineering College, V V Nagar
safvan465@gmail.com

## ABSTRACT

*T There are plenty of security software in market; each claiming the best, still we daily face problem of viruses and other malicious activities. If we know the basic working principal of such malware then we can very easily prevent most of them even without security software. Hackers and crackers are experts in psychology to manipulate people into giving them access or the information necessary to get access. This paper discusses the inner working of such attacks. Case study of Spyware is provided. In this case study, we got 100% success using social engineering techniques for deception on Linux operating system, which is considered as the most secure operating system. Few basic principal of defend, for the individual as well as for the organization, are discussed here, which will prevent most of such attack if followed.*

## 1. INTRODUCTION

We are living in the Internet world, and we heard daily regarding virus and hackers. We all install antivirus and anti-Spyware software but still the virus infects our system. Today our every business is linked with IT systems. All major banks provides embanking, we purchase ticket and do shopping online. In our country still the IT penetration in our day-to-day life is not that much compared to other like USA, Europe countries. Our businesses are linked with IT and hence with computers. Computers get hacked by hackers or infected by the Virus, Worms and that affect businesses to great extend.

### 1.1. Impact of Malware Activity

Intensions behind Hacking and Malware are different, their threats also vary. Hacking threat can be for financial gain or personal revenge. Spyware threat can be social or personal. Virus threat can be economical. Bitnet and Trojan ware threat leads to social and national security. There are various types of threats
1. Personal Threat / Organizational Threat
2. National Security Threat
3. Economical Threat
Most Parasites writer or phishers have their main purpose as Money and hence most of the time it is related to banking. Brand attacked in November 2009 were banking, ecommerce, IT services and other.

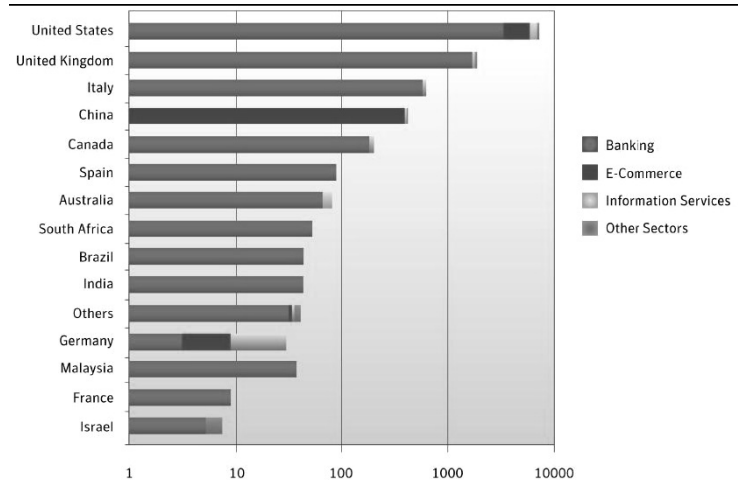

[State of Phishing: Monthly Report: December 2009 by Symantec]

There were 29 countries whose brands were attacked. During November 2009, In China, the e-commerce sector remains a primary target. Due the 2010 FIFA World cup, Phishers are launching attacks masquerading World cup related sites. From the above report you can see that most of the time phishers are attacking banking system.

### 1.2 Anti Virus and Anti Spyware Systems

As we think, hacker and Malware activity as technical problem, we always search for the better anti-Spyware or anti-virus software. Our anti-virus or anti-Spyware monitors running application but they do not check for the people problem. Usually people do not know the basic working of how malware penetrate into the system? What cause they can do? How to prevent them?

## 2 PSYCHOLOGY

According to the study of psychology, human being has nature to be helpful when people are in real need, the tendency to trust people, the fear of getting into trouble and tries to escape from it, get something free or without doing much of work. Hacker and crackers tries to attack this technique hence people need to be trained to defend against it.

### 2.1. Introduction to Social Engineering

It is a process of deceiving people into giving away access or confidential information, is a formidable threat to most secured networks. It is an art of persuasion. Social Engineering techniques and philosophies behind it is very old [11]. The story of the Trojan horse, made famous by the Greek epic poet Homer in The Odyssey. It was one of the most ingenious social engineering tricks in the history of humankind. Edwards named it after the social engineering technique used by the Greeks [11]. This attack is the most powerful attack as no hardware or software can prevent it or even can not defend it. This attack deals with Psychology and hence people need to be trained to defend against it. Followings are the few definition of Social Engineering by some authors.
1. "…the art and science of getting people to comply with your wishes."[2]
2. "Social Engineering - A euphemism for non-technical or low-technology means - such as lies, impersonation, tricks, bribes, blackmail, and threat - used to attack information systems." [2]
Each Social Engineering attacks are different and unique but they do have some common pattern. That pattern has four phases (Information Gathering, Relationship Development,

Exploitation and Execution). Social Engineering attack and/or may even incorporate the use of other more traditional attack techniques to achieve the desired end result. We see that social engineering not only is a serious threat but also the inherent human weakness for misperception of human mind to guess the true motive of the persuader.

In the Information gathering phase, Target victim or victims are identified. Once the target is identified, Next step is to identify various luring section which will excite the victim. Example way old "I Love You" virus was working on this principal, as every one of us wants to know who love me. Once information is gathered regarding victim and his interest area. Next step is we try to find or establish the relationship with victim. Here in "I Love You" virus, if virus is coming for some friend then you get more excited. After that attack is planned with already built software or new Spyware will be created. Now everything is set. The last step of the attack is execution

## 2.2. Types Social Engineering skills

Following are the few Skills to exploits user to get access to your system.

1. Impersonating staff: This is an art of inventing scenario to persuade a target to release information or perform an action and is usually done through email or telephone. Most powerful and danger trick for gaining physical access to the system is to pretend to be someone from inside the company. Users gave their password to a "stranger" on a phone call to a member of the IT staff. This is especially true if the caller implies that their account may be disabled and that they might not be able to get important e-mail or access needed network shares if they don't cooperate [3]. It is the most time consuming attack as it requires research to get information regarding target to establish the legitimacy in the mind of target.

2. Playing on users' sympathy the social engineer may pretend to be a worker from outside, perhaps from the phone company or the company's Internet service provider [2]. Nature of people is to help a person who's in trouble.

3. Intimidation tactics social engineers may need to turn to stronger stuff: intimidation. In this case, the social engineer pretends to be someone important -- a big boss from headquarters, a top client of the company, an inspector from the government, or someone else who can strike fear into the heart of regular employees. He or she comes storming in, or calls the victim up, already yelling and angry. [2] They may threaten to fire the employee they don't get the information they want.

4. Hoaxing: A hoax is an attempt to trick the people into believing something false is real. It also may lead to sudden decisions being taken due to fear of an untoward incident.

5. Creating confusion: Another ploy involves first creating a problem and then taking advantage of it. It can be as simple as setting off a fire alarm so that everyone will vacate the area quickly, without locking down their computers. Social engineers can then use a logged-on session to do their dirty work. [2]

6. Dumpster diving: Someone from the company throwing away junk mail or routine mail / letter of the company without ripping the document. If the mail contained personal information, or credit card offers that dumpster diver could use to carry out identity theft. Dumpster diver also searches for information like company organization chart, who reports to whom, especially management level employee who can be impersonated to hack important detail. Dumpster diving information can be used in impersonation attack.

7. Reverse social Engineering: An even sneakier method of social engineering occurs when a social engineer gets others to ask him or her questions instead of questioning them. These social engineers usually have to do a lot of planning to pull it off, placing themselves in a position of seeming authority or expertise.

8. Mail: The use of an interesting subject line triggers an emotion that leads to accidental participation from the target. There are two common forms. The first involves malicious code; this code is usually hidden within a file attached to an email. The intention is that an unsuspecting user will click/open the file; for example, 'I Love You' virus, 'Anna Kournikova' worm.

9. A phishing technique that has received substantial publicity of late is "vishing," or voice phishing. Vishing can work in two different ways. In one version of the scam, the consumer receives an e-mail designed in the same way as a phishing e-mail, usually indicating that there is a problem with the account. Instead of providing a fraudulent link to click on, the e-mail provides a customer service number that the client must call and is then prompted to "log in" using account numbers and passwords. The other version of the scam is to call consumers directly and tell them that they must call the fraudulent customer service number immediately in order to protect their account. Vishing criminals may also even establish a false sense of security in the consumer by "confirming" personal information that they have on file, such as a full name, address or credit card number [4]. Vishing actually emulates a typical bank protocol in which banks encourage clients to call and authenticate information [5].

## 3 CASE STUDY

As Social Engineering is the most powerful attack, we tried to check how effective the social engineering on the Linux. Linux is considered as most secure operating system but as we have discussed even the most secure system can be break by targeting weak link (people). Following case study shows the impact of social engineering if plugged with Spyware.

### 3.1. Implementation of Case Study

We have created a Spyware for the Linux which logs the information typed by the user in Linux environment. We have not put the Spyware on the wild means in real environment but to get statistics related to Spyware with social engineering tactics, we tried to achieve it three ways and all these techniques are based on social engineering [6].

**Case-1**: Enthusiasm of fun: As for this attack, first we gather information like we used the person whom we know. We gathered the information like he uses the Operating system as Linux; He is fond of Linux shell script programming. Second stage is relation ship development which is already established as we choose person who trust us. We sent it to friend who uses Linux as desktop operating system in mail. Subject line of the mail was "Shell Script for Fun". As frequently you got mail from your friend having attachment, you open it as it pretend to be from friend and safe. You get trapped because even you can send mail with any fake name using open mail relay SMTP servers. This is psychological strategy and it is customized attack, as we have chosen the individual. Some parasite writers use customize approach for some specific victim while some uses general approach to trap unknown victims and if that technique get successful then many people will get trap in it.

**Case-2**: Eagerness to know great thing: Second case we gone for same principal first information gathering, relation establishment and then deception. We choose persons who are fond of hacking and cracking activity. Case-2 targets such user who loves hacking and cracking. Even on internet, if you search for free tools for hacking and cracking, you will get it for free. But many of such software itself hack your system. On internet: I have put link of this shell script with Name "Tool to hack in Windows" to friend. And they clicked on it, downloaded it and ran it.

**Case-3**: Hoaxing: As people think, Linux is secure than windows but don't know by what percentage and they want information on this aspect. So fake Linux report containing the Shell Script as the case. As a news to friend – Linux security report. As people normally follows the link as it is report on Linux. As all this techniques uses social engineering, human get trapped. Our spyware requires the root privileged.

Following table shows the result of above techniques. [6]

| Case Study Result | | |
|---|---|---|
| Case | Number of target | Success |
| 1 | 5 | 80 |
| One instance didn't work, as run at office in Underprivileged user mode and hence didn't work. | | |
| 2 | 3 | 100 |
| Worked 100 percent as every one has executed it in root mode. | | |
| 3 | 2 | 50 |
| One of the target users has not followed the Link provided in mail | | |

This Spyware doesn't exploit any of the Linux vulnerability but uses social engineering to attack people and result shows that it may get worst if deliver in the wild with few more techniques [6]. This Spyware gets more effective because Linux, itself provide very powerful tool.

### 3.2. Other Example of Social Engineering Attacks

Social Engineering is used by hackers and crackers to hack the target machine or to spread virus and Malware application. To introduce the Social Engineering, we have to give some real example, which can be understood easily by the non-technical person.

Following are the few attacks for spreading Spyware

1. Piggybacked software installation: User is lured to install the software for free and with that software automatically some Spyware get installed which will monitor and even tamper your data. That software might be claiming of game or media player or any useful software [1].

2. Mail: you get mail from your friend or from unknown mail id with some interesting or alert subject line like "Hey check your machine" or "You might be infected" you open it and you get infected [1].

3. Fake anti Spyware: There are various utilities claiming anti-Spyware but actually they are Spyware or some application enticed with hacking tool but actually hacking your system. The weakest security link, which concerns playing with human psychology to get the confidential details out of him by appearing to be 'genuine and concerned'

4. Spam mail claiming "You won the lottery" or claiming to be selling some genuine medicine for good result. This all are the social engineering to lure the target to provide some information which can be used to gain financial or social or personal gain.

## 4 PREVENTION

User education is the first and most powerful defense against social engineering, backed up by strong, clear (written) policies that define when and to whom (if ever) users are permitted to give their passwords, open up the server room, etc. Strict procedures should be laid down. By implementing authentication system (smart cards/tokens or, even better, biometrics), you can thwart a high percentage of social engineering attempts. Even if the social engineer manages to learn the password, it will be useless without the second authentication factor. A successful defense against the social engineering depends on having good policies in place ensuring that all employees follow them Social engineering attacks are most powerful attacks as the defense against it is not the software system but the people which in themselves quite unpredictable. Still using few counter measures we can prevent some of the attacks.

**Following are the prevention techniques for personal defense.**

1. We have to be suspicious of any e-mail with urgent requests for personal financial information or threats of termination of online accounts.

2. Unless the e-mail is digitally signed, you can't be sure it wasn't forged or "spoofed." because any one can mail it by any name hence when it is stating some important better to check for the full headers.

3. Phishers typically ask for information such as usernames, passwords, credit card numbers, social security numbers, etc. and such information normally won't be asked by the genuine organization online.

4. Phisher e-mails are typically not personalized, while valid messages from your bank or e-commerce company generally are. "Phisher e-mails start some thing like "Dear customer" but there are some attacks which are customized or more advance which uses your personal information and if the attack is specifically for you then it will be customize like our case study.

5. When contacting your financial institution, use only channels that you know from independent sources. (e.g., information on your bank card, hard-copy correspondence, or monthly account statement), and don't rely on links contained in e-mails, even if the sites looks genuine.

6. Always ensure that you're using a secure website when submitting credit card or other sensitive information via your Web browser. Check in the address bar URL must start with https:// instead of http://

7. Regularly log into your online accounts and change password frequently.

8. Regularly check your bank, credit and debit card statements to ensure that all transactions are legitimate.

9. Don't assume that you can correctly identify a website as legitimate just by looking at its general appearance.

10. Avoid filling out forms in e-mail messages or pop-up windows that ask for personal financial information because it might be used by spammers as well as phisher for future attack.

**Following are the few counter measures for the organizational institute.**

1. Well defined and documented security policy: In this process company set the standards and guidelines form the foundation of a good security [7].

2. Acceptable usage Policy: for acceptable business usage of email, computer system, telephone and network as well as other hardware like pen-drive.

3. Personnel security: A screening of prospective employees, contractor to ensure that they do not pose a security threat to the organization [3].

4. Information Access Control: Password usage and guidelines for generating password, access authorization and accounting procedure, installation procedure. Automated password reset and synchronization tools can lift the responsibility of managing password from tech support and help desk without placing an undo burden on end user [3].

5. Protection from Malware like Spyware, virus, adware, Trojan etc using software systems. Like firewalls, antispyware and anti-virus software with regular updating of patches. These will ensure filtering of major security breach incidents [3].

6. Awareness and Education: Giving education to the user about the common techniques employed and used by the social engineer is an important part of security system. For example, a knowledgeable user can be advised that he/she should never give out any information without the appropriate authorization and that he/she should report any suspicious behavior[9][10].

A good training and awareness program focusing on the type of behavior required will undoubtedly pay for itself. By providing real incident example, social engineering can be implemented effectively in the system.

7. Audits and compliance: Policy gets effective only when it gets implemented and everyone conforms to the policy. Hence auditing the usage and make sure everyone compliance to the rules [9] [10].

8. Security Incident Management: When a social engineering attacks occurs make sure service desk staff knows how to manage such attack. As each attack is different, system will get new data and hence its need to be manages for future use. Hence reporting and storage of such incident should be done properly [10].

**Followings are the few points of email usage policy given by SANS institute.**
1. Email system shall not to be used for the creation or distribution of any disruptive or offensive messages or even forwarded message. Employees who receive any emails with this content from any <COMPANY NAME> employee should report the matter to their supervisor immediately.
2. Using a reasonable amount of <COMPANY NAME> resources for personal emails is acceptable, but nonwork related email shall be saved in a separate folder from work related email.
3. Employees shall have no expectation of privacy in anything they store, send or receive on the company's email system. <COMPANY NAME> may monitor messages without prior notice. <COMPANY NAME> is not obliged to monitor email messages.
4. Any employee found to have violated this policy may be subject to disciplinary action, up to and including termination of employment.

## 5 CONCLUSION

We might have the most secure network or clear policies still as humans are unpredicted due curiosity and greed without concern for the consequences, we could face our own version of a Trojan tragedy [11]. A paradox of social engineering attacks is that people are not only the biggest problem and security risk, but also the best tool to defend against these attacks. Organizations must fight social engineering attacks by establishing policies and procedures that define roles and responsibilities for all users and not just security personnel. As well as organization must ensure that, these policies and procedure are executed by users properly hence regular training needs to be given on the latest such incidents.

## REFERENCES

[1] Malware : Threat to the Economy, Survey Study by Mosin Hasan, National Conference IT and Business Intelligence (ITBI - 08)

[2] White paper: Avoiding Social Engineering and Phishing Attacks,Cyber Security Tip ST04-014, by Mindi McDowell,Carnegie Mellon University, June 2007.

[3] Book of 'People Hacking' by Harl

[4] FCAC Cautions Consumers About New "Vishing" Scam, Financial Consumer Agency of Canada, July 25, 2006.

[5] Schulman, Jay. Voice-over-IP Scams Set to Grow, VoIP News, July 21, 2006.

[6] Spying Linux: Consequences, Technique and Prevention by Mosin Hasan, IEEE International Advance Computing Conference (IACC'09)

[7] Redmon,- audit and policy Social Engineering manipulating source , Author: Jared Kee,SANS institute.

[8] White paper 'Management Update: How Businesses Can Defend against Social Engineering Attacks' published on March 16, 2005 by Gartner.

[9] White paper, Social Engineering:An attack vector most intricate to tackle by Ashish Thapar.

[10] The Origin of Social Engineering Bt Heip Dand MacAFEE Security Journal, Fall 2008.

[11] Psychology: A Precious Security Tool by Yves Lafrance,SANS Institute,2004.

[12] SOCIAL ENGINEERING: A MEANS TO VIOLATE A COMPUTER SYSTEM, By Malcolm Allen, SANS Institute, 2007

[13] Inside Spyware – Techniques, Remedies and Cure by Mosin hasan Emerging Trends in Computer Technology National Conference.